\documentstyle[epsfig]{aipproc}

\def\teq#1{$\, #1\,$}                         
\def\erg{\varepsilon}         
\def\split{\gamma\to\gamma\gamma}               
{\catcode`\@=11                                                                 
\gdef\SchlangeUnter#1#2{\lower2pt\vbox{\baselineskip 0pt \lineskip0pt           
  \ialign{$\m@th#1\hfil##\hfil$\crcr#2\crcr\sim\crcr}}}}                        
\def\gtrsim{\mathrel{\mathpalette\SchlangeUnter>}}

\def\pmb#1{\setbox0=\hbox{#1}%
  \kern-0.0125em\copy0\kern-\wd0
  \kern0.025em\copy0\kern-\wd0
  \kern-0.0125em\raise0.0433em\box0 }
\def\vol#1#2{$\;$ {\bf #1}, #2}                         

\def\apj{{\it Ap. J.}}   \def\apjl{{\it Ap. J. (Lett.)}}
\def\apjs{{\it Ap. J. Suppl.}}

\def\prd{{\it Phys. Rev. D}}  
\begin{document}
\title{A New Class of Radio Quiet Pulsars}

\author{Matthew G. Baring\thanks{Compton Fellow, Universities Space
        Research Association} and Alice K. Harding}
\address{Laboratory for High Energy Astrophysics, Code 661,\\
         NASA Goddard Space Flight Center, Greenbelt, MD 20771}

\maketitle

\begin{abstract}
The complete absence of radio pulsars with periods exceeding a few
seconds has lead to the popular notion of the existence of a high $P$
\it death line\rm .  In the standard picture, beyond this boundary,
pulsars with low spin rates cannot accelerate particles above the
stellar surface to high enough energies to initiated pair cascades
through curvature radiation, and the pair creation needed for radio
emission is strongly suppressed.  In this paper we postulate the
existence of another pulsar ``death line,'' corresponding to high
magnetic fields $B$ in the upper portion of the $\dot{P}$--$P$ diagram,
a domain where few radio pulsars are observed.  The origin of this high
$B$ boundary, which occurs when $B$ becomes comparable to or exceeds
$10^{13}$ Gauss, is again due to the suppression of magnetic pair
creation $\gamma\to e^+e^-$, but in this instance, primarily because of
ineffective competition with the exotic QED process of magnetic photon
splitting.
This paper describes the origin, shape and position of the new ``death
line,'' above which pulsars are expected to be radio quiet, but perhaps
still X-ray and $\gamma$-ray bright.
\end{abstract}

\section*{Introduction}

Due to the broad range of period derivatives observed for isolated
radio pulsars, the population spans over four decades in their
estimated spin-down field strengths (e.g. Taylor, Manchester and Lyne
1993).  However, none have inferred (dipolar) fields exceeding a few
time \teq{10^{13}}Gauss, suggesting that there is an observational bias
against observing high-field pulsars.  This bias could be due to a
complete absence of neutron stars with fields much above
\teq{10^{13}}Gauss, or perhaps radio emission is somehow suppressed at
such high field strengths, diminishing their observability.  The former
hypothesis has no intrinsic theoretical basis, and is contradicted by
the suggestion (Duncan and Thompson 1992) that soft gamma repeaters
have supercritical fields, above \teq{10^{14}}Gauss.  Hence, it is of
interest to examine the latter possibility, that high field pulsars do
not produce radio emission; this is the focus of this paper.

Magnetic one-photon pair production, \teq{\gamma\to e^+e^-}, has
traditionally been the only gamma-ray attenuation mechanism assumed to
operate in polar cap models for radio (e.g. Sturrock, 1971) and
gamma-ray pulsars (Daugherty \& Harding 1982, 1996; Sturner \& Dermer
1994), providing the means for both types of pulsars to radiate
efficiently.  Such an interaction can be prolific at pulsar field
strengths, specifically when the photons move at a substantial angle
\teq{\theta_{\rm kB}} to the local magnetic field.  Pair creation has a
threshold of \teq{2m_ec^2/\sin\theta_{\rm kB}}.  The exotic
higher-order QED process of the splitting of photons in two,
\teq{\split}, will also operate in the high field regions near pulsar
polar caps and until very recently, has not been included in polar cap
model calculations.  Magnetic photon splitting has recently become of
interest in neutron star models of soft gamma repeaters (Baring 1995),
and Harding, Baring and Gonthier (1997) have determined that splitting
will play a prominent role in the formation of spectra for PSR1509-58,
the gamma-ray pulsar having the lowest high-energy spectral turnover,
around $\sim 1$ MeV.

The key property of photon splitting that renders it relevant to
neutron star environs is that it has {\it no} threshold, and can
therefore attenuate photons below the threshold for pair production,
\teq{\gamma\to e^+e^-}.  Hence, when it becomes comparable to
\teq{\gamma\to e^+e^-}, it will diminish the production of secondary
electrons and positrons in pair cascades.  Since pairs are probably
essential to the generation of radio emission (e.g. Sturrock 1971),
such a ``quenching'' of pair creation can potentially provide a pulsar
``death-line'' at high field strengths; this phenomenon is the subject
of this paper.  While about a dozen radio pulsars have spin-down
magnetic fields above $10^{13}$ Gauss, little attention was paid to
\teq{\split} in pulsar contexts prior to the launch of the Compton
Gamma-Ray Observatory (CGRO) in 1991 because until then, the three
known gamma-ray pulsars had estimated field strengths of less than a
few times \teq{10^{12}}Gauss.  The detection of PSR1509-58 by the OSSE
and COMPTEL experiments on CGRO provided the impetus to focus on
high-field neutron star systems.

\section*{Quenching of Pair Creation in Pulsars}

In polar cap models, pair cascades in radio and gamma-ray pulsars are
initiated by relativistic electrons either via curvature radiation
(e.g. Daugherty and Harding 1982), or by their resonant (magnetic)
Compton scattering (e.g. Sturner and Dermer 1994) with thermal X-rays
that emanate from the stellar surface.   The cascades are perpetuated
and amplified by synchrotron radiation interspersed with generations of
pair creation.  The nature of these three processes is well
understood.  For relativistic electrons with Lorentz factor
\teq{\gamma}, photons produced by these mechanisms are collimated to
angles \teq{\sim 1/\gamma} to the direction of the electron's
momentum.  Furthermore, the produced radiation in each of these
processes is highly polarized.  The degree of polarization \teq{{\cal
P}} of synchrotron and curvature emission (they are identical: see, for
example, Jackson 1975) is \teq{(p+1)/(p+7/3)} for power-law electrons
\teq{n_e(\gamma)\propto\gamma^{-p}}, and is in the 60\%--70\% range
(e.g. see Bekefi 1966), favouring the production of photons in the
\teq{\perp} state.  Here the label \teq{\parallel} refers to the state
with the photon's \it electric \rm field vector parallel to the plane
containing the magnetic field and the photon's momentum vector, while
\teq{\perp} denotes the photon's electric field vector being normal to
this plane.  Likewise, it can be deduced from Herold's (1979)
expression for resonant Compton scattering in the Thomson limit that
the upscattered photons are predominantly in the \teq{\perp} state
also, achieving \teq{{\cal P}\sim}50\%.  Hence any pair creation
in pulsars is primarily initiated by photons with polarization state
\teq{\perp}, thereby simplifying the considerations here.

In this paper we will assess how effective photon splitting is relative
to magnetic pair creation in attenuating photons that are produced by
these radiation mechanisms.  In doing this, we propagated photons
outwards from some point on or above the stellar surface, computing
their attenuation probabilities for both of these processes.  Of
specific interest is the {\it escape energy}, \teq{\erg_{\rm esc}}, the
energy below which (for each mechanism) photons escape the neutron star
magnetosphere without attenuation.  We fully include the general
relativistic effects of a Schwarzschild spacetime, and details of the
geometry and propagation set-up are described at length in Harding,
Baring and Gonthier (1997).  In that work, which focused on the
high-field test-case pulsar PSR1509-58, it was clearly demonstrated
that pair creation is suppressed when photon splitting dominates it at
higher field strengths.  Also, generally, larger polar cap
sizes favour the suppression of cascades and hence radio emission.

We computed the magnetic fields \teq{B_{\rm d}} for given polar cap
angles (colatitudes) \teq{\Theta} for which the escape energies for
splitting and pair creation were equal, so that for \teq{B\gtrsim
B_{\rm d}} pair creation is strongly suppressed by splitting.  Since
the multiplicity of pairs in a pulsar cascade rapidly becomes large in
just a few generations, quenching is extremely abrupt and effective at
high fields.  Therefore, we expect a rapid decline in pair creation and
hence also radio luminosity when \teq{B} rises above \teq{B_{\rm d}}.
Remembering that the polar cap size is coupled to the pulsar period
\teq{P} (in flat space time \teq{\Theta\approx  (2\pi/P)^{1/2}\,
(R_{\rm ns}/c)^{1/2}}, and we included general relativistic corrections
to this), the so-defined (\teq{B}, \teq{\Theta}) relationship becomes a
critical curve on the \teq{P}-\teq{\dot{P}} diagram.  This curve
delineates the phase spaces for radio-loud and radio-quiet pulsars, and
examples are depicted in Figure~1, along with the latest population
distribution from the Princeton pulsar catalogue.  This boundary
delineates a zone where pair creation is suppressed, like its long
period counterpart.  However, there is no pulsar evolution across the
boundary (without field evolution): high field pulsars are born
radio-quiet. Hence it is not a true death-line, just a border to the
radio quiet region.

\begin{figure}                                          
\centerline{\epsfig{file=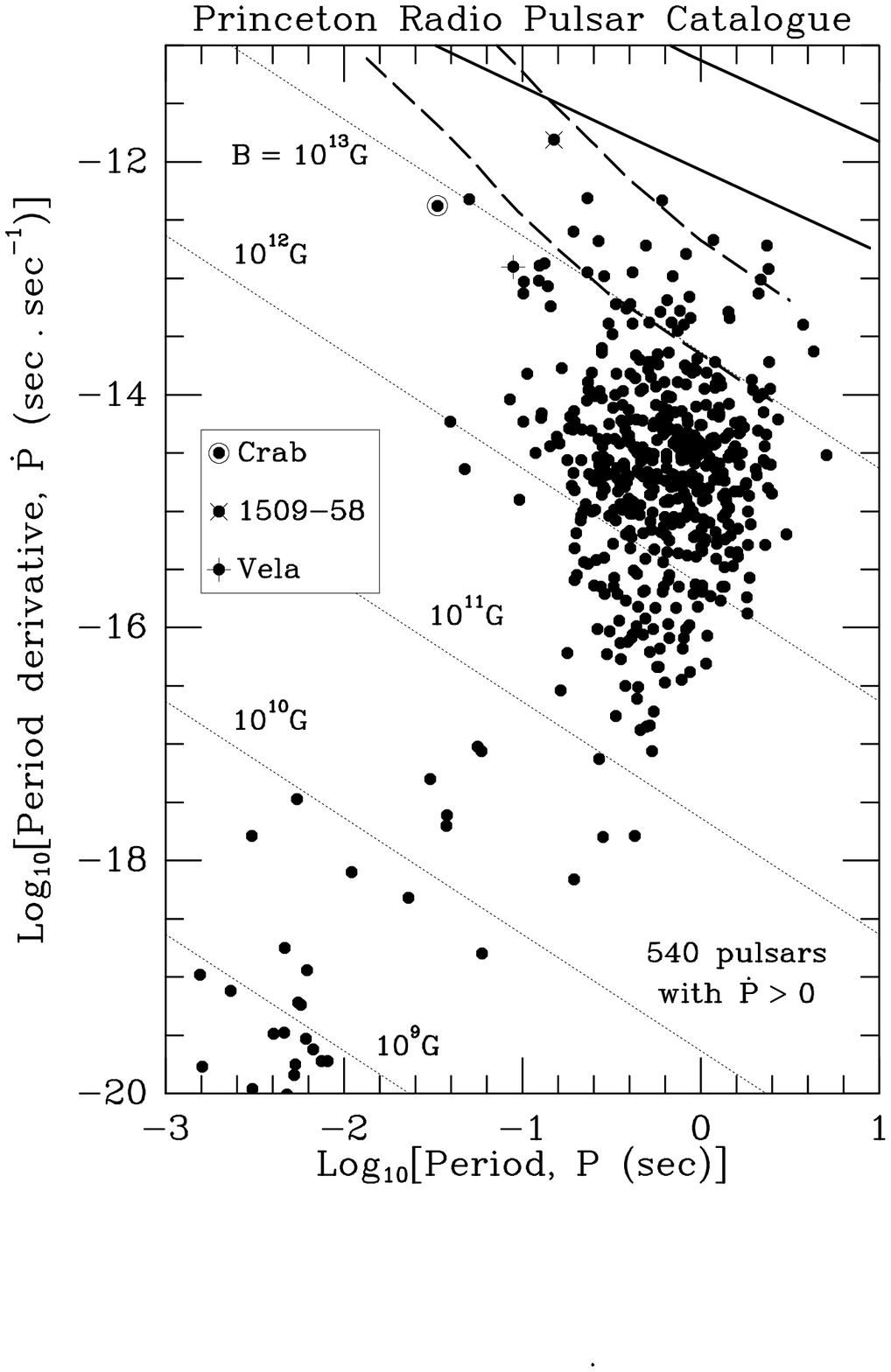,height=7.5in}}
\vspace{-2.7truecm}
\caption{The $P$--$\dot{P}$ diagram for the latest Princeton Radio
Pulsar catalogue (May 3, 1995: see also Taylor, Manchester and Lyne
1993) together with four possible high-field ``death'' lines (heavy
solid and dashed curves), above which pulsars are radio-quiet.  For the
four cases depicted, the solid curves represent situations where
photons that seed potential cascades (e.g. curvature radiation) are
initially beamed very close to the field lines, while the dashed curves
have such photons initially propagating at an angle of $0.57^\circ$ to
the local field.  For both these scenarios, the lower curves are for
emission from the stellar surface, and the upper ones are for photons
originating half a stellar radius above the surface.  The light dotted
diagonal lines define contours of constant $B$, as labelled, and three
gamma-ray pulsars in the diagram are marked as indicated in the inset.}
\label{fig:ppdot}
\end{figure}

There are four examples of such boundaries in Figure~1 because the
results differ according to the initial angles of photons with respect
to {\rm B}, and their original location.  For photons that start out
almost along the field, as in a curvature radiation-initiated cascades
(the solid curves), \teq{B_{\rm d}\propto\Theta^{-1/3}\;\Rightarrow\;
\dot{P}\propto P^{-5/6}}.  When photons initially have appreciable
angles to {\bf B}, as can be the case in resonant Compton-initiated
(IC) cascades,  photon splitting competes more effectively with pair
creation for smaller polar cap sizes and the ``death-line'' drops to
lower field strengths.  Clearly the position of the line, which marks
{\it surface} fields, strongly depends on the radius of photon origin
since the physics of this problem couples to the magnitude of \teq{B}.
Hence there is, at present, significant uncertainty in the location of
the radio-quiet boundary, principally because the location of the
acceleration of primary electrons is not fully understood.  Note that
ground state pair creation also becomes prevalent for high \teq{B}
(Harding and Daugherty 1983), thereby aiding cascade cessation and
lowering the radio-quiet boundary in the \teq{P}-\teq{\dot{P}}
diagram.  Note also that there is marginal evidence for a drop in
pulsar radio-luminosity when their fields exceed around \teq{3\times
10^{13}}Gauss, contrary to the slow increase with \teq{B} seen for
lower spin-down fields.

Clearly, when pair creation is suppressed and pulsars become radio
quiet, they can still emit $\gamma$-rays prolifically, via the
primary electrons and spectral reprocessing via splitting.  Hence it is
reasonable to conjecture that the radio quiet pulsars may actually be a
class of objects formerly known as Gemingas.  Motivations for searching
for such $\gamma$-ray pulsars are therefore self-evident.  Soft
$\gamma$-ray observability is perhaps governed by \teq{(B/P^2)^{1/2}},
favouring sources to the upper left of the \teq{P}-\teq{\dot{P}}
diagram, which should guide pulsar searches with OSSE and COMPTEL.  On
the other hand, hard $\gamma$-ray (i.e. EGRET, {\teq{>100}MeV)
observability implies no spectral cutoffs (as in PSR1509-58), and
favours small \teq{\Theta}, in the upper right of the diagram (for high
\teq{B}).  The very shapes of the ``death-lines'' in Figure~1 indicate
that radio searches for high-field pulsars should focus on sub-second
periods and high \teq{\dot{P}}.  In conclusion, in the polar cap model,
the physics described here may well imply the existence of a
radio-quiet pulsar population with high surface fields.

\end{document}